\documentclass{ws-p10x7}

\begin{document}

\title{Quark Gluon Plasma - Recent Advances}

\author{Grazyna Odyniec}

\address{Lawrence Berkeley National Laboratory, Berkeley, CA 94720,
USA\\E-mail: G\_Odyniec@lbl.gov}

\twocolumn[\maketitle\abstract{
While heavy ion collisions at the SPS have produced excited strongly interacting matter near the conditions for quark deconfinement, the RHIC may be the first machine capable of creating quark-antiquark plasmas sufficiently long-lived to allow deep penetration into the new phase. A comprehensive experimental program addressing this exciting physics has been put into place. 
Presented here are preliminary results from Au+Au at $\sqrt{S}$ = 130 GeV obtained during the first RHIC run and some CERN SPS results from Pb+Pb at $\sqrt{S}$ = 17 GeV (particularly relevant to QGP search).
}]

\section{Introduction}%1

The most popular among conventional scenarios of the beginning of the Universe is the Big Bang model. Since George Gamow first proposed it in 1948, the idea of an explosive birth has steadily and successfully battled competing theories. A number of researchers have refined the model over the intervening decades. Using general relativity and some basic physical laws, the model as it exists today, envisages a beginning from an extremely small, hot, dense initial state some 20 billion years ago. At time zero, the Universe is believed to have been all energy. Almost instantly, free quarks and gluons started to condense out of the rapidly expanding energy cloud. After a few microseconds, quarks began to coalesce into mesons and nucleons. Only minutes later, protons and neutrons began to form nuclei, and the evolution of stars and galaxies was launched. 

Unlike other areas of physics, the study of the birth and evolution of the Universe has the statistics of only one event to rely on. One can, however, attempt to reverse this process and create in the laboratory a ``fireball'' like the one at the beginning of the Universe by colliding heavy nuclei at velocities near the speed of light. Such collisions offer the unique opportunity to compress nuclear matter to very high density (several times its normal density) and to heat it to very high temperatures. If collisions are energetic enough and the energy density reaches its critical value for phase transition (about 1 GeV/fm$^3$ according to the lattice-QCD\cite{FK}), protons, neutrons, and other nuclear ingredients might separate into quarks and gluons forming a quark gluon plasma state (``Little Bang''). 

The CERN SPS heavy ion program ($\sqrt{S}$ = 17 GeV) was designed with this in mind (in the experiment the volume and energy density are controlled by the size of the colliding nuclei and their collision energy). 

And in fact, the analysis of the central Pb+Pb collisions at CERN SPS energies indicates that matter with an energy density of several GeV/fm$^3$ was created at the early stage of the collisions,\cite{QM95} exceeding significantly the critical energy density.  The transient existence of a quark-gluon plasma in the collisions, if produced, is expected to modify the evolution of the system compared to a scenario of confined hadronic matter. 
Indeed, some results from analysis of Pb+Pb data fall into a common pattern that could not be explained within a hadronic scenario, but may very well signal the onset of an expected QCD phase transition. The higher energies would provide larger energy density and, probably, would amplify these effects. Therefore, the first results from heavy ion collider experiments at much higher energy are awaited with great anticipation.

The direct comparison between experiment and theory is neither simple nor straightforward for both theoretical and experimental reasons. On the theory side, one needs to remember that basic assumptions of QCD are not fully satisfied by the experiment. The majority of experimental difficulties arise from the fact that detectors look at the collisions after freeze-out. 

This overview starts with a short presentation of aspects of the SPS Pb+Pb collisions, possibly related to the QGP. It will be followed by a rather broad discussion of  first (often preliminary) results from experiments underway at the Relativistic Heavy Ion Collider (RHIC, $\sqrt{S}=$ 130 and 200 GeV) in the USA, and others which are planned to start in 2005 with the Large Hadron Collider (LHC, $\sqrt{S}=$ 5.5 TeV) at CERN. 

\section{Departure from Hadronic Scenario at CERN SPS}%1

The recent CERN SPS results from Pb on Pb collisions at $\sqrt{S}$ = 17 GeV have confirmed,\cite{QM99} beyond doubt, previous ``hints'' of new physics emerging from the analysis of S+S interactions at 200 GeV/c. The most significant surprises, including strangeness enhancement (particularly multi-strange hyperons), anomalous J/$\psi$ suppression, and change in the spectral shape of e$^+$e$^-$ mass distributions in the vector meson domain, are discussed briefly in the next sections.

\subsection{Strangeness Production}\label{subsec:wpp}%1.1

Strangeness production has long been suggested as a probe of the central fireball region due to the possibility of unique production mechanisms should a QGP be formed.\cite{RM,KMR} The main arguments discussed in the literature, are related to (1) a much lower threshold for production of $s\bar{s}$ pairs in a QGP compared to an hadron gas, (2) contribution from gluons abundantly present in the plasma via the fusion process (gg$\rightarrow$ $s\bar{s}$), (3) Pauli blocking where at large baryon density production of $s\bar{s}$ pairs might be favored if the lowest available u,d quark levels are larger than 2m$_s$, and (4) compatibility of the equilibration time in the plasma with the time needed for two nuclei to traverse each other at this energy (for hadronic gas the equilibrium time is about 20-30 times longer).
For a detailed discussion see reference \cite{GO}. 
 
The first measurement of strangeness enhancement at SPS energy comes from the NA35/NA49 large acceptance spectrometer experiment.\cite{GA,BA} 

% fig 1 strangeness enhancement
\begin{figure}%1
\epsfxsize190pt
\figurebox{}{}{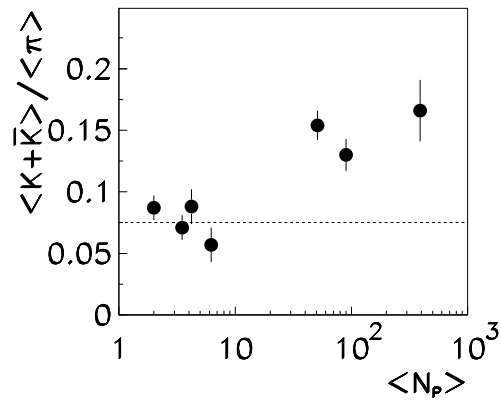}
\caption{Multiplicity ratio $\langle K \rangle /\langle \pi \rangle$ in full phase space for pp, pA and AA collisions plotted versus the average number of participating nucleons. See text for details.}
\label{fig:radk}
\end{figure}

Figure 1 shows the K/$\pi$ ratio ($\sim$ strangeness/entropy, where 
kaons account for about 70 $\%$ of the overall strangeness production) 
for pp, pA and AA collisions as a function of centrality ($\approx$ number of participating nucleons). A global enhancement factor of about 2 is observed in central A+A collisions relative to pA (and pp). More interestingly, the enhancement is of the same size in all three reactions: S+S, S+Ag and Pb+Pb suggesting that the system already reached some kind of saturation in S+S collisions. This rules out the interpretation of strangeness enhancement as a consequence of hadronic re-interactions.\cite{GO}

The more differential analysis reveals a new aspect of charged kaon behavior: the energy dependence of strangeness (actually strangeness per entropy as it is shown in Fig.2 and 3) is significantly different in the case of K$^-$/$\pi^-$ and K$^+$/$\pi^+$. The K$^-$/$\pi^-$ ratio rises semi-monotonically with the energy, whereas the K$^+$/$\pi^+$ rises sharply until $\sqrt{S}$ reaches $\sim$ 5-10 GeV, flattens out, and then decreases. 

% Figure 2: K+/Pi+ vs sqrt S
\begin{figure}%1
\epsfxsize190pt
\figurebox{}{}{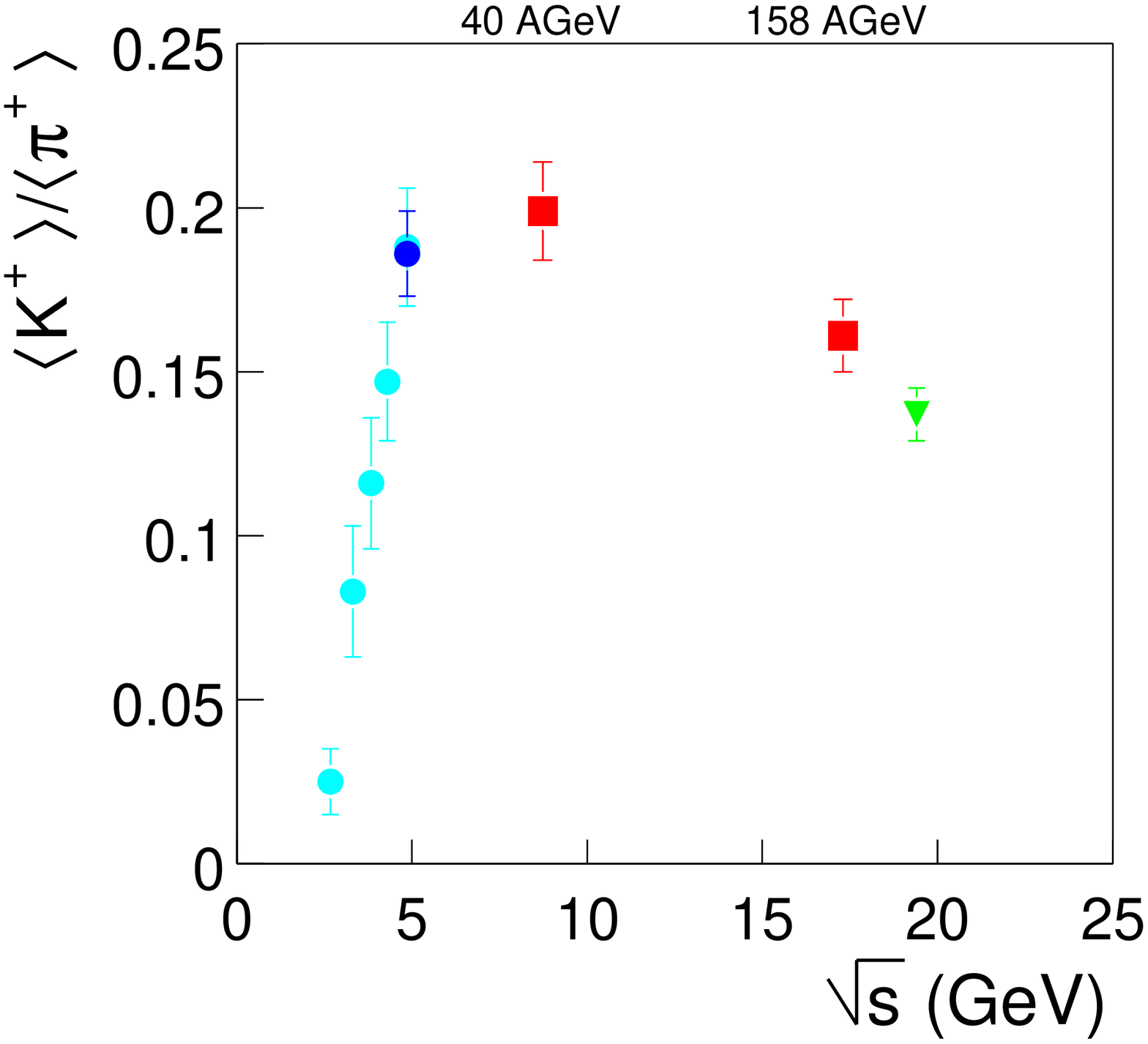}
\caption{Energy dependence of the $\langle K^+ \rangle /\langle \pi^+ \rangle$ ratio.}
\label{fig:radk}
\end{figure}

% Figure 3: K-/pi- vs sqrt S
\begin{figure}%1
\epsfxsize190pt
\figurebox{}{}{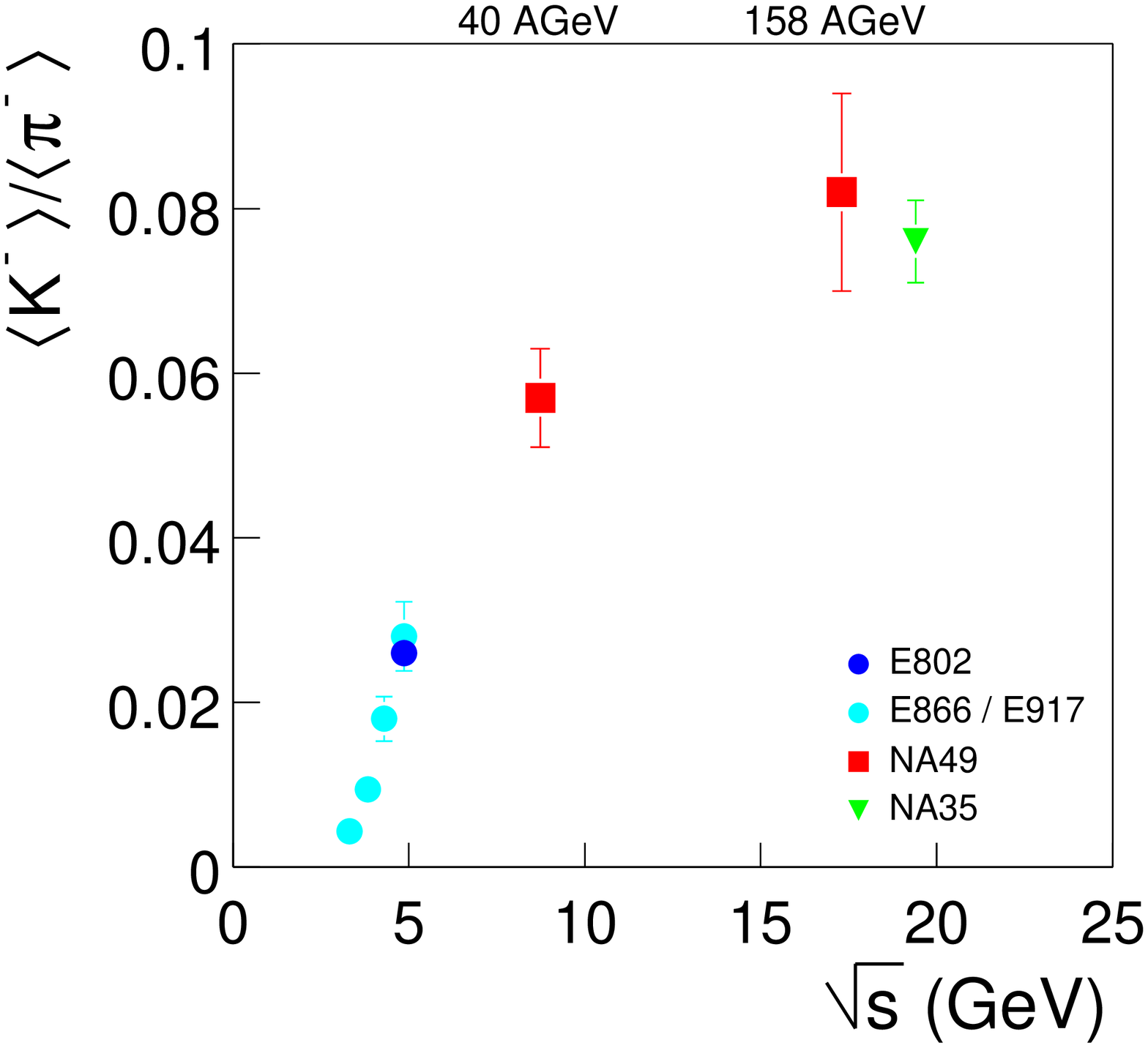}
\caption{Energy dependence of the $\langle K^- \rangle /\langle \pi^- \rangle$ ratio.}
\label{fig:radk}
\end{figure}

Since K$^+$ are produced in the associated production with $\Lambda$, this measurement is sensitive to the baryon density in the interaction volume. The K$^+$/$\pi^+$ maximum at $\sim$ 40 GeV/c reflects the highest baryon density and, perhaps, the most favorable ``sweet spot'' for the phase transition. Further experiments at CERN, with 20 and 30 GeV/c planned for next year, may provide more of an explanation.

The precise measurements of strange hyperons and their anti-particles show a systematic increase of their yields with respect to p+Pb ($\Lambda$, $\Xi$, $\Omega$ WA97/NA57 data\cite{Fi,Vi}) and with respect to pp ($\Xi^-$ NA49 data\cite{Ap}). In Fig.4 the measured multiplicities of strange particles per participant is shown, using p+Be results as a reference.

% Fig 4 multistrange
\begin{figure*}
\epsfxsize30pc
\figurebox{16pc}{32pc}{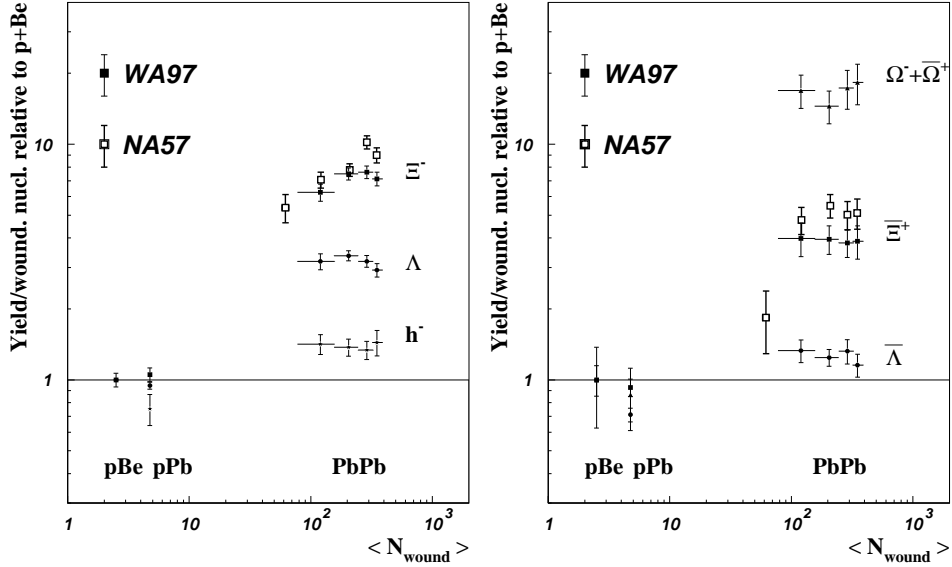}
\caption{Negative hadrons and hyperons yields per participant, normalized to p+Be results, measured by WA97/NA57. The yields in Pb+Pb are above what one would expect if they were proportional to the number of wounded nucleons, represented by the straight line. Note, that there is an enhanced production even in the case of negative hadrons.
\label{fig:radish}}
\end{figure*}

 The hierarchy of enhancements foreseen in case of QGP formation\cite{RM} is clearly observed, with an $\Omega$ enhancement factor of about 17. The other remarkable result is that for N$_{part}\geq$ 100 a saturation of enhancement is observed. In order to determine the onset point of this phenomenon, the preliminary NA57 data (available so far only for $\Xi$ hyperons) which covers the region of peripheral Pb+Pb collisions is used. The open squares in Fig.4 mark the NA57 data points. The slight discrepancy between WA97 and NA57 reflects the magnitude of systematic errors. While the results of NA57 are still preliminary, they already  suggest that the observed saturation breaks down with decreasing centrality. Thus, it appears to be characteristic of a particular class of very central events. The nearly complete saturation, together with the fact that the value of chemical freeze-out temperature extracted within the framework of the statistical model\cite{BM,Be} is so close to the one predicted for the phase transition, suggests that the system quite probably crosses the phase space boundary shortly before the chemical freeze-out point and that observed saturation comes essentially from the partonic phase.    

The study of the transverse mass spectra has been carried out by several experiments (NA49, NA44, WA97). It was soon realized that the inverse slope T of the transverse mass distribution, representing the temperature of the system at decoupling time, increases steadily with the mass of the considered particle (see Fig.5).

%  fig 5  T slopes, omega departure
\begin{figure}%1
\epsfxsize190pt
\figurebox{}{}{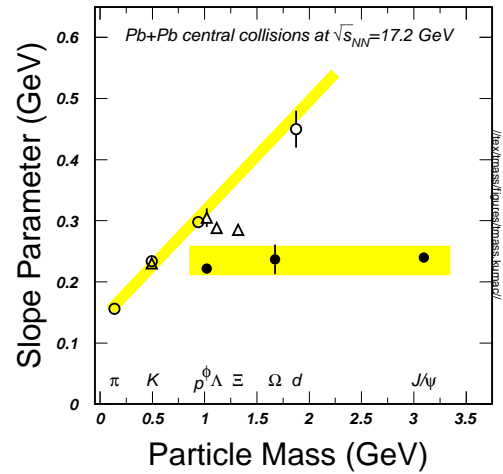}
\caption{Inverse slope parameter T of the transverse mass distribution as a function of the mass of particles in Pb+Pb collisions at $\sqrt{S}$ = 17 GeV.}
\label{fig:radk}
\end{figure}

 This was explained as arising from the relativistic superposition of the expanding system plus a radial collective flow, caused by the explosion of the ``fireball.''\cite{Bea,App} From the outside of this ``Little Bang'' it looks like a blue shift of the temperature, conceptually similar to the red-shift observed for the background radiation from the Big Bang.   
The departure of the $\Omega$ from the systematics might be due to an early freeze-out  ($\Omega$'s can not form resonances with pions and therefore may decouple very early from the expanding gas). This would indicate that the enhancement of $\Omega$ does not originate from the late stages of the reaction, but rather is due to the pre-hadronic phase.\cite{Hec}

\subsection{J/$\psi$ Suppression}\label{subsec:wpp}%1.1

Another signal of quark-gluon plasma formation is the expected suppression of J/$\psi$ production in high energy nuclear collisions.\cite{Ma} At the early stage of the collision, suppression of the J/$\psi$ and $\psi$' resonances is expected due to a QCD mechanism, analogous to Debye screening in a QED plasma, or from interaction of the charmonia state with the hard gluons present in deconfined matter. It has been demonstrated that neither J/$\psi$ nor $\psi$' could be broken up by a hadronic co-moving environment, because hard processes should in general suffer no effect by the later stage of the heavy ion collision.\cite{Sa} Therefore, J/$\psi$ and $\psi$' suppression, if observed, would provide a relatively ``clean'' signature of a QCD phase transition. 

The NA38 data from several proton-nucleus and nucleus-nucleus collisions shows the J/$\psi$ meson suppression in nuclear collisions, as compared to the reference process of Drell-Yan (DY) pair production. In p-A and A-A collisions, up to central S+U, the trend of J/$\psi$ suppression is very well described by the absorption process of its pre-resonant state.\cite{Ka} The NA50 measurements of J/$\psi$ production in central Pb+Pb show a significant change of pattern, manifested by the abrupt onset of a much stronger suppression for impact parameters smaller than 8-8.5 fm.

The anomalous suppression in Pb+Pb (see Fig.6) collisions appears as a sharp discontinuity from the nuclear absorption mechanism, and keeps increasing with centrality. 

% Fig 6 - J/psi
\begin{figure}%1
\epsfxsize190pt
\figurebox{}{}{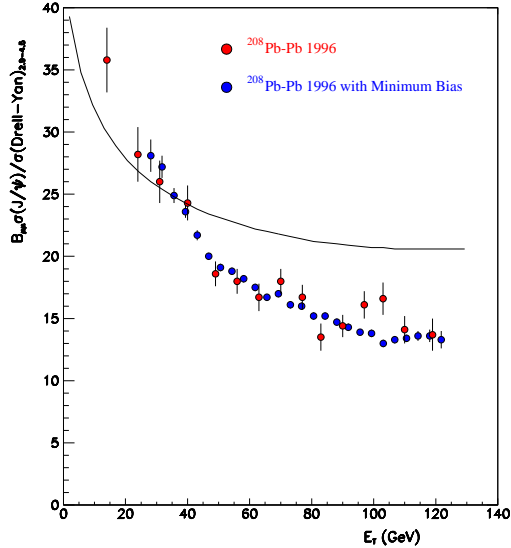}
\caption{Ratio of the $J/\psi$ over Drell Yan as a function of the transverse energy E$_t$ in Pb+Pb at $\sqrt{S}$ = 17 GeV. The full line represents the absorption model. See text for details.}
\label{fig:radk}
\end{figure}

The full line represents the absorption model with a cross section for dissociation of $c\bar{c}$ pairs by a nucleon of about 6.2 mb.
The observed anomalous degree of suppression, far beyond the $\sim$6 mb of pre-hadronization break-up, was tentatively interpreted as a result of creation of the energy density in the core of the Pb+Pb interaction volume, comfortably sufficient for creation of a partonic, rather than a hadronic state.\cite{Kl}

\subsection{Low-mass Dilepton Production}\label{subsec:wpp}%1.1

Leptons are produced during the entire evolution of the collision. Since they do not re-interact (the mean free path for electromagnetic interactions is much larger than the size of the interaction volume) leptons are assumed to provide information on all subsequent stages of the system evolution. A careful analysis might be able to unfold the entire space-time history of the collision and possibly separate the contributions from the partonic and hadronic phases.

The NA45/CERES experiment studied the mass spectra of inclusive e$^+$e$^-$ pairs in p+A (Fig.7) and S+Au (Fig.8) at 200 GeV/c and in Pb+Au at 158 GeV/c.

% fig 7 dileptons
\begin{figure}%1
\epsfxsize190pt
\figurebox{}{}{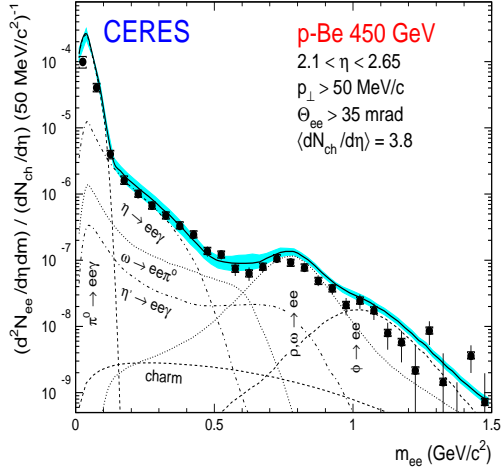}
\caption{Mass spectra of inclusive $e^+e^-$ pairs in 450 GeV p+Be collisions showing the data (full circles) and various contributions from hadronic decays. Systematic (brackets) and statistical (bars) errors are plotted independently of each other.}
\label{fig:radk}
\end{figure}

% fig 8 dileptons
\begin{figure}%1
\epsfxsize190pt
\figurebox{}{}{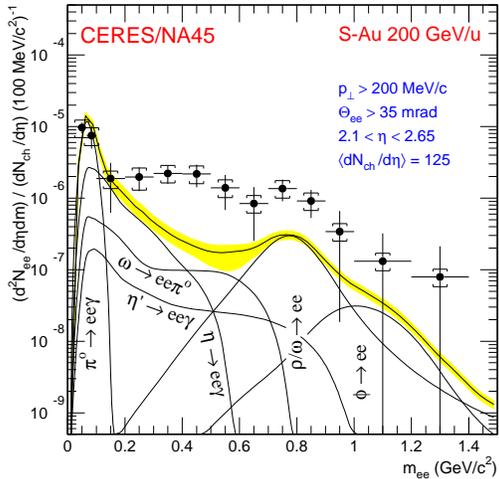}
\caption{Mass spectra of inclusive $e^+e^-$ pairs in Pb+Pb collisions at 200 GeV/c showing the data (full circles) and various contributions from hadronic decays. Systematic (brackets) and statistical (bars) errors are plotted independently of each other.}
\label{fig:radk}
\end{figure}

 A significant excess of low-mass electron pairs is seen in S+Au collisions, over the so-called ``cocktail'' of known hadronic sources (Dalitz decays of $\pi^0$, $\eta$, $\eta$' and decays of the $\rho$, $\omega$ and $\phi$ resonances) estimated by scaling from pp collisions. In the case of proton induced interactions, the low-mass spectra are, within errors, well explained by the electron pairs from hadronic decays.
The enhancement observed in Pb+Pb is very similar in shape to the one in S+Au.
The excess in the low-mass spectrum was also reported in the muon channel by the HELIOS/3 experiment.
Those results have triggered a wealth of theoretical activity. There is consensus that a simple superposition of pp collision can not explain the data.\cite{Ag1} The pion annihilation channel ($\pi^+\pi^-\rightarrow\rho\rightarrow$ $l^+l^-$), added to the ``cocktail'', improved to some degree the agreement with A+A data. This was tentatively interpreted as first evidence of thermal radiation from the dense hadronic matter formed in the collision. The quantitative agreement with the data required further processes (not present in pp and pA) to be included in the calculations. The number of models with different underlying scenarios both partonic and hadronic were developed.\cite{LKB,Ra,RCW,VK,Hu} It turns out that all calculations led to similar results because in fact the dilepton production rates calculated via hadronic and partonic models are very similar at SPS conditions.\cite{Ra}  

The present uncertainties in the data are still too large to convincingly argue for necessity of in-medium meson mass treatment that might be interpreted as a signal of the onset of hadronic chiral restoration mass loss in high density hadronic medium.

\subsection{Lessons learned from CERN SPS}\label{subsec:wpp}%1.1

Studies of Pb+Pb interactions at CERN energies have demonstrated that the critical value for the QCD phase transition was reached and exceeded. Did one get to see the first glimpses of a new phase ?

Perhaps ... No definite answer yet. 

\section{RHIC Expectations}%1

15 years of the heavy ion physics program at CERN SPS resulted in several suggestive results, but there was no clear discovery made. The newest machine, the Relativistic Heavy Ion Collider (RHIC) at Brookhaven National Laboratory designed to run with heavy ion beams up to gold nuclei at energy of $\sqrt{S}$=200 GeV, is expected to further exceed the critical energy density, making a transition to a QGP state more feasible.\cite{QM99}
Heavy ion collisions at RHIC energies will explore regions of energy and particle density that are significantly beyond those reachable at SPS. The energy density of thermalized matter created at RHIC is estimated to be 70$\%$ (at $\sqrt{S}$=130 GeV) higher than at SPS, implying a much greater initial temperature, $T_0$, or higher multiplicity, or both. Due to the higher initial parton density, thermalization would also happen more rapidly. As a consequence, the ratio of the quark-gluon plasma lifetime to the thermalization time would increase substantially. This implies that the ``fireballs'' created in heavy ion collisions at RHIC will spend a large fraction of their lifetime in a purely partonic state. Thus, the time window available for experiments to probe a new phase widens significantly.
 
The full set of reference data (pp, pA with many ion species), will complete the nucleus-nucleus program. The analysis of pp and pA interactions, in addition to providing the reference for AA, will allow one to study nuclear parton distribution functions.  
 
\section{First Year Results from Au+Au Collisions at $\sqrt{S}$=130 GeV}%

RHIC began operation in June 2000 with Au beams at $\sqrt{S}$=130 GeV (total $\sqrt{S}$= 25 TeV) extending the available center of mass energy in nucleus-nucleus collisions by nearly a factor of 8 over SPS ($\sqrt{S}$=17 GeV). The four heavy ion experiments: STAR (TPC based large acceptance spectrometer), PHENIX (small acceptance, multi-arm spectrometer), PHOBOS (a Si-based, compact multiparticle spectrometer) and BRAHMS (a forward and midrapidity hadron spectrometer) are designed and built to allow a comprehensive study of heavy ion collisions at RHIC energies. 
They take various approaches to search for deconfinement phase transition to QGP. The STAR experiment concentrates on measurements of hadron production over a large solid angle in order to analyze single- and multi-particle spectra and to study global observables on an event-by-event basis. The PHENIX experiment is focused on measurements of lepton and photon production and has capability of measuring hadrons in a limited range of pseudorapidity. The two smaller experiments BRAHMS and PHOBOS are focused on single- and multi-particle spectra.
All four performed well during the first run, and took high quality data.\cite{QM01}
The very first physics results are already published,\cite{QM01,S01} with some of them being presented in the next chapters. During the second run, one year later (2001), the collider reached the design energy value. While the 12 week run is still in progress, the very preliminary analysis of $\sqrt{S}$=200 GeV data is already under discussion. 

\subsection{Multiplicity}\label{subsec:wpp}%1.

The first measurements indicated an increase of about 70$\%$ in the charged multiplicity for central collisions compared to previous measurements.\cite{PH}
Figure 9 shows the corrected, normalized multiplicity distribution for minimum bias Au+Au collisions  within $\mid \eta \mid\leq$0.5 and p$_t\geq$100 MeV/c taken with the STAR TPC detector.\cite{Ac}
The STAR TPC resides in a solenoidal magnet operated at 0.25 T for the data shown on Fig.9.
The data were normalized assuming a total hadronic inelastic cross section of 7.2 b for Au+Au at $\sqrt{S}$= 130 GeV, derived from Glauber model calculations. The shape of the h$^-$ minimum bias multiplicity distribution is dominated over much of the range by the nucleus-nucleus collision geometry, consistent with findings at lower energies, whereas the tail region of the spectrum is determined by fluctuations and acceptance. The systematic error on the vertical scale is estimated to be 10$\%$ and is dominated by uncertainties in the total hadronic cross section and the relative contribution of the first bin. The systematic error on the horizontal scale is 6$\%$ for the entire range of multiplicity and is depicted by horizontal error bars on a few data points only. These overall futures are also observed in the Hijing calculations\cite{Hij}. The gray area represents the 5$\%$ most central collisions (360 mb). The normalized pseudorapidity distribution of those events within $\mid \eta \mid \leq$1, both for p$_t\geq$100 MeV/c and for all p$_t$ is displayed in Fig.10. The error bars indicate the uncorrelated systematic errors, the statistical errors are negligible. The correlated systematic error applied to the overall normalization is estimated to be smaller than 7$\%$. The $\eta$ distribution is almost constant within $\mid\eta\mid\leq$1 as expected from a boost invariant source. The dashed line shows the data published by the PHOBOS Collaboration\cite{PH} for the average  charge multiplicity measured over the range $\mid \eta \mid \leq$1. There is good agreement between the measurements for the average multiplicities. 
It should be noted that in Pb+Pb at $\sqrt{S}$=17 GeV the pseudorapidity distribution of charged hadrons\cite{Ag} and rapidity distribution of negative hadrons (assuming the pion mass)\cite{NA49} were found to peak at midrapidity, signaling a significant change in the longitudinal phase space distribution between the SPS and RHIC. Pseudorapidity distributions like the one in Fig.10 should constrain some of the important ingredients used in model calculations, particularly the initial gluon distributions and possibly the evolution in the early phase of the collisions, both of which are expected to significantly influence particle production.
The data published by three other experiments (PHOBOS, PHENIX and BRAHMS)\cite{QM01} for the average charged multiplicity measured in the same rapidity range do agree within 10$\%$ with the STAR measurement. This consistency is a quite impressive and very encouraging.
%
% fig 9 star multiplicity
\begin{figure}%1
\epsfxsize190pt
\figurebox{}{}{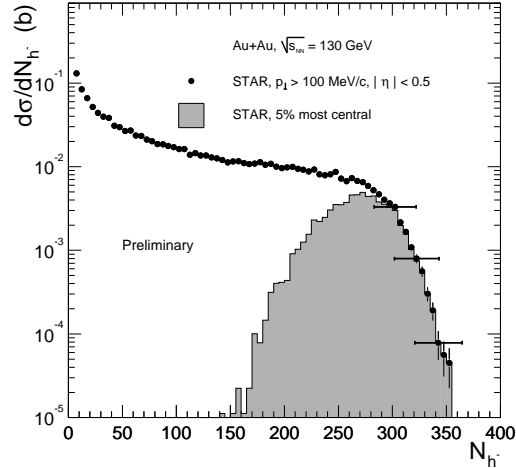}
\caption{Multiplicity distributions of negative hadrons at mid-rapidity measured in minimum bias Au+Au collisions at $\sqrt{S}$ = 130 GeV in STAR TPC acceptance (p$_t\geq$ 100 MeV/c, $\mid \eta \mid \leq$ 0.5). The gray area represents the 5$\%$ most central collisions. A solid curve depicts the predictions of HIJING calculations.}
\label{fig:radk}
\end{figure}
%
%
%
% fig 10  pseudorapidity - very centrals
\begin{figure}%1
\epsfxsize190pt
\figurebox{}{}{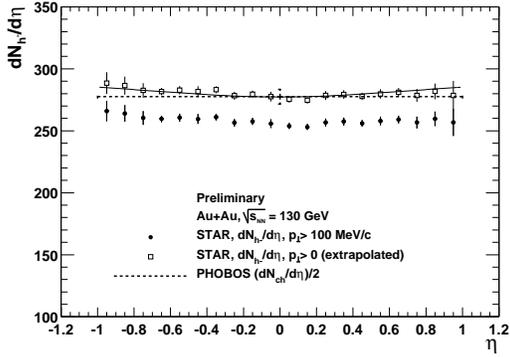}
\caption{Pseudorapidity distributions for negative hadrons measured in the most central 5$\%$ of Au+Au at $\sqrt{S}$ = 130 GeV collisions for p$_t\geq$ 0.1 GeV/c (solid dots) and extrapolated to p$_t$ = 0 (open squares). The dash line shows the data published by the PHOBOS Collaboration, see text for details.}
\label{fig:radk}
\end{figure}

Fig.11 shows the comparison of charged particle density per participating pair versus the c.m. energy. Both points at RHIC energy ($\sqrt{S}$ = 65 GeV and $\sqrt{S}$ = 130 GeV) taken by the PHOBOS collaboration for the 6$\%$ most central Au+Au are compared to pp and $p\bar{p}$ data (open symbols) and to the NA49 Pb+Pb (central 5$\%$) data (filled square). The energy dependence of the charged hadron multiplicity is shown to be enhanced in heavy ion collisions relative to $pp$ and $p\bar{p}$ reactions. The model calculations (solid and dashed lines in Fig.11) are discussed in \cite{WG1}.   

%fig 11 XNwang pict
\begin{figure}%1
\epsfxsize190pt
\figurebox{}{}{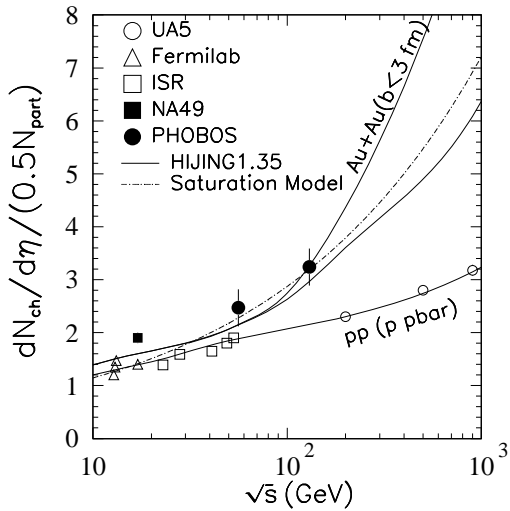}
\caption{Charged particle rapidity density per participating baryon pair versus the c.m. energy. The PHOBOS data (filled circles) for Au+Au at RHIC energy and the NA49 Pb+Pb data (filled square) are compared to $pp$ and $p\bar{p}$ points (open symbols). See text for details.}
\label{fig:radk}
\end{figure}

\subsection{Baryons at Midrapidity - $\bar{p}/p$ ratio}\label{subsec:wpp}%1.

The degree of baryon stopping (or baryon number transport) affects many stages of the dynamical evolution  of the collision: initial parton equilibrium, particle production, thermal and/or chemical equilibrium and the development of collective expansion. Experimentally, the information on baryon stopping can be accessed though measurements of antiproton to proton yield. The study of $\bar{p}/p$ shows that at $\sqrt{S}$=130 GeV the ratio at midrapidity is significantly smaller ($\bar{p}/p$ = 0.65 $\pm$ 0.01(stat.error) $\pm$ 0.07(syst.error) for the minimum bias collisions) than 1, indicating an overall excess of protons over antiprotons (no net-baryon free midrapidity yet). On the other hand, it is dramatically increased over the AGS value ($\bar{p}/p$ = 0.00025 $\pm$ 10$\%$)\cite{Ah} and SPS value ($\bar{p}/p$ = 0.07 $\pm$ 10$\%$)\cite{Si,Ka}.  
This is presented in Fig.12 where the $\bar{p}/p$ ratio for central heavy ion collisions is shown as a function of the center of mass energy, $\sqrt{S}$.

% fig 12 - p_bar/p ratio vs S
\begin{figure}%1
\epsfxsize190pt
\figurebox{}{}{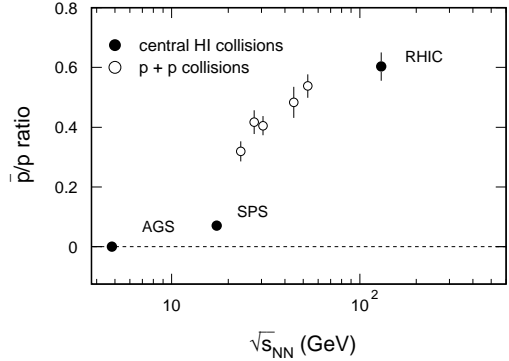}
\caption{Ratio of $\bar{p}/p$ in midrapidity for RHIC at $\sqrt{S}$ = 130 GeV is compared with AGS ($\sqrt{S}$ = 5 GeV) and SPS ($\sqrt{S}$ = 17 GeV) values. Open symbols mark p + p corresponding points.}
\label{fig:radk}
\end{figure}

 Interestingly, the value of the $\bar{p}/p$ ratios in p+p collisions\cite{Ro,AB} (open symbols) appear to be very close to RHIC values.

The comparison of RHIC ratios to heavy ion results at lower energies indicate that while the midrapidity $\bar{p}/p$ ratio increases significantly with the collision energy, there is still a substantial excess of baryons over antibaryons present at midrapidity at $\sqrt{S}$=130 GeV.

\subsection{Chemical Freeze-Out}\label{subsec:wpp}%1.

Recently, much of the theoretical effort has been devoted to the analysis of particle production within the frame of the statistical model (both in the case of elementary and heavy ion collisions). Data indicate that matter at freeze-out may be described by equilibrium distributions, i.e. particle ratios are well fitted with only two parameters: temperature T and chemical potential $\mu_B$.\cite{Be,Bec,He,Cl,PBM}

This pattern appears to hold also at RHIC energies. The measurements of midrapidity ratios of particles ($\pi, K^0, p, \phi, \Lambda, \Xi$) and their antiparticles agree well with the statistical model with T$\approx$ 170-190 MeV and $\mu_{B}\approx$ 40-50 MeV\cite{Xu} - see Figure 13.  

% fig 13 - masashi + nu
\begin{figure}%1
\epsfxsize190pt
\figurebox{}{}{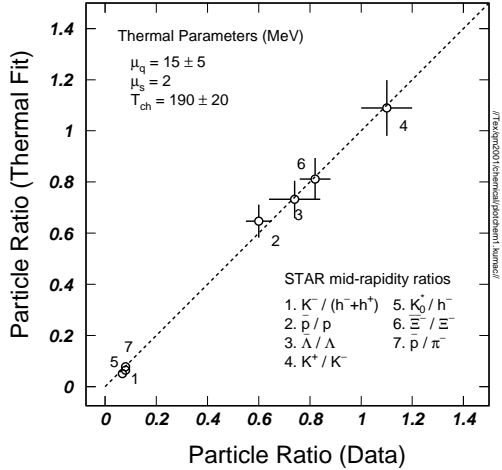}
\caption{STAR preliminary particle ratios versus results of thermal model fits. For the thermal parameters of temperature and chemical potential, see text.}
\label{fig:radk}
\end{figure}

Note, that to make a comparison to a thermal model requires the ratios to be measured over the entire acceptance. So far, only midrapidity ratios are available. On a very preliminary level, with the assumption of a boost invariant scenario, one may speculate that these data already indicate that there are enough final state interactions to drive the system towards equilibrium. 

Taking further the implications of the thermal model, one can try to place the RHIC point on the phase space diagram (T vs $\mu_B$) at chemical freeze-out - Fig.14.\cite{Cl}

% fig 14  phase space nu
\begin{figure}%1
\epsfxsize195pt
\figurebox{}{}{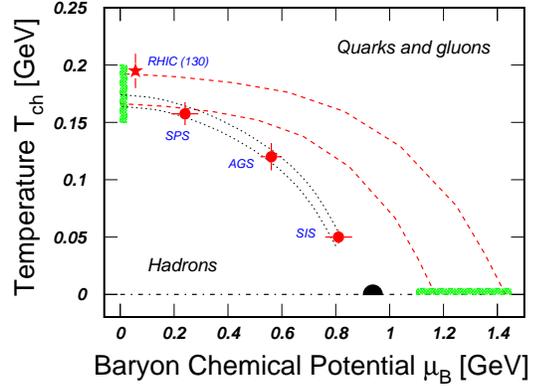}
\caption{ Phase space plot T versus $\mu_B$. Dashed-line represents the boundary between interactions involving hadronic and partonic degrees of freedom. Dotted-lines represent the model described in the reference 42. 
The ground state of nuclei is shown as half circle.}
\label{fig:radk}
\end{figure}

 At $\sqrt{S}$=130 GeV the system seems to already be in the quark-gluon domain. The $\sqrt{S}$=200 GeV measurements will become available soon. If both consistently stay above the phase transition boundary, will one be able to claim the discovery of a global, thermalized system? Certainly not yet, because (1) comparison/test with thermal models should be carried out with data measured in full acceptance and (2) all thermal models do rely on implicit assumption of equilibration.

\subsection{Early Stage of Collisions - Hard Probes}\label{subsec:wpp}%1.

During the early stage of the collision, high mass and high momentum objects are created. The ``hard scale'', characterizing these probes, is the squared momentum transfer, Q$^2$, necessary for their production. Studies of the quark gluon plasma can be done efficiently by using ``hard probes'' e.g. high $p_t$ jets or photons, heavy quarkonia, and W$^\pm$ or Z$^0$ mesons. So far only jet production (leading particle approximation\cite{Wa}) is addressed with the first year RHIC data.\cite{QM01}  

High p$_t$ quark and gluon jets, due to their hard production scales, materialize very early during the collision and are thus embedded into and propagate through the dense environment of the ``fireball'' as it forms and evolves. In particular, they are expected to suffer a loss of energy as they traverse the dense medium created in the collision zone. The loss of energy is supposed to be proportional to energy density, therefore through this interaction, they measure the property of the environment and are sensitive to the formation of a QGP.\cite{Wa,Gy} Large transverse momentum probes are easily isolated experimentally from the background of soft particles produced in the collision. The high $p_t$ of the probes ensures that the medium effects are perturbatively calculable, which strengthens their usefulness as quantitative diagnostic tools. Fig.15 shows the transverse momentum distribution of negatively charged hadrons for the 5$\%$ most central Au+Au collisions (STAR).

% fig 15  transverse momentum star
\begin{figure}%1
\epsfxsize190pt
\figurebox{}{}{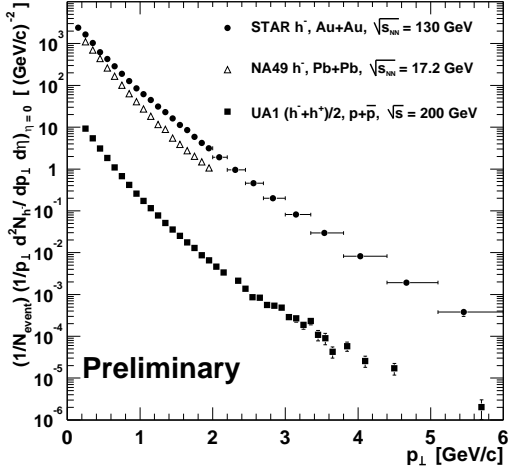}
\caption{Transverse momentum distributions for hadrons at mid-rapidity in central Au+Au collisions at RHIC (STAR preliminary), central Pb+Pb in NA49, and $p+\bar{p}$ in UA1.}
\label{fig:radk}
\end{figure}

 For comparison, the data from central collisions of Pb+Pb measured at midrapidity at $\sqrt{S}$=17 GeV by the NA49 collaboration\cite{HA} and p$_t$ distribution of an average of the positive plus negative hadrons from $p\bar{p}$ at $\sqrt{S}$=200 GeV are displayed.\cite{Al} 
All three distributions follow a simple power-law, but the spectrum from STAR is flatter than the other two. Fig.16 shows the ratio of the STAR spectrum to a smooth parameterization of the UA1 spectrum, scaled appropriately for the energy and geometry difference,\cite{Du} as a function of p$_t$. 

% fig 16  star comp to ua1
\begin{figure}%1
\epsfxsize190pt
\figurebox{}{}{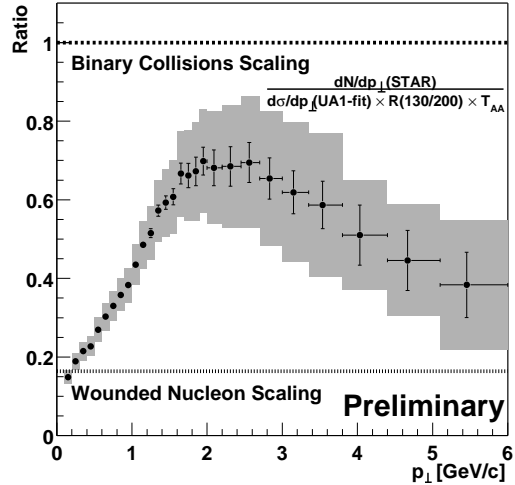}
\caption{Ratio of STAR negative hadron $p_t$ distribution from Au+Au collisions with the scaled UA1 average hadron $p_t$ distribution in $p+\bar{p}$ collisions. Vertical bars denote error on measurement, gray boxes cumulative error including error on UA1 scaling.}
\label{fig:radk}
\end{figure}

The ratio at low p$_t$ is well reproduced by a scaling with the number of wounded nucleons in the collision, and it rises with higher p$_t$. It reaches a maximum at about p$_t$=2 GeV/c, then it decreases signaling a suppression of hadron yields at high p$_t$ in Au+Au relative to the $p\bar{p}$ reference. Hadron yields at high p$_t$ (in absence of nuclear effects) are expected to scale as the number of binary collisions. STAR data, lying significantly below the binary collision limit, indicates that fragmentation products of hard-scattered partons are suppressed in Au+Au collisions. It is tempting to assume that the first sign of the presence of partonic energy loss in nuclear collision was detected.\cite{GP,WG}
A similar observation is reported by the PHENIX collaboration in respect to the charged hadrons and $\pi^0$ yields in central (10$\%$ of most central events) and peripheral (60-80$\%$ of the geometrical cross section) collisions.\cite{PHEN} The neutral pions are measured by two separate detectors: a lead-scintillator (PbSc) sampling calorimeter and lead-glass \v{C}erenkov (PbGl) calorimeter. The two analysis have very different systematics and Fig.17 shows the agreement of their final $\pi^0$ spectra. The data are compared to the binary scaled yield from N+N collisions. Above $p_t\sim$ 2 GeV/c, the spectra from peripheral collisions appear to be consistent (within systematic errors) with a sum of underlying N+N collisions. The spectra from central collisions, in contrast, are systematically below the scaled N+N expectation, both when compared to p+p and to Au+Au peripheral collisions.

While the hadron suppression in central collisions, measured by both the STAR and PHENIX experiments, is in qualitative agreement with the predictions of energy loss by partons traversing a dense medium, other explanations can not be eliminated yet. Further measurements to reduce large systematic errors ($\sim30\%$ presently), together with new, more accurate pp (not $p\bar{p}$) and pA reference data are needed before a definite conclusion can be made.

% phenix figure 17
\begin{figure}%1
\epsfxsize210pt
\figurebox{}{}{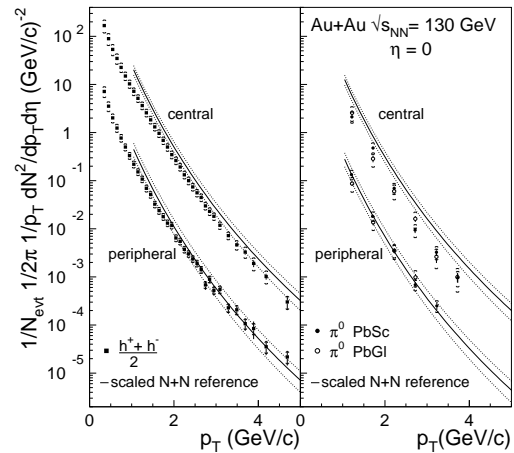}
\caption{The yields per event at mid-rapidity for charged hadrons (left) and neutral pions (right) are shown as a function of p$_t$ for 60-80$\%$ (lower) and 0-10$\%$ (upper) event samples, with the $\pi^0$ results from the PbSc and PbGl analyses plotted separately. The error bars indicate the statistical errors on the yield; the surrounding brackets indicate the systematic errors. Shown for reference are the yields per collision in N+N, of charged hadrons and neutral pions, respectively, scaled with number of collisions. The bands indicate both the uncertainties in N+N reference and in the scaling factor.}
\label{fig:radk}
\end{figure}

\subsection{Azimuthal Anisotropy}\label{subsec:wpp}%1. 
   
The complementary method for addressing the early evolution of the system uses the analysis of the azimuthal anisotropy of the transverse momentum distribution (elliptic flow) for non-central collisions. The anisotropic emission of particles ``in'' and ``out'' of the reaction plane (defined by the beam and the impact parameter directions) is characterized by the second harmonic Fourier coefficient, $v_2$, of the azimuthal distribution of particles with respect to the reaction plane. Details of the $v_2$ dependence on beam energy and centrality are thought to be sensitive to the phase transition between confined and deconfined matter.\cite{Ko,Ol}
The peak elliptic flow measured at RHIC energy by the STAR collaboration reaches 6$\%$,\cite{Fl} while values at AGS and SPS are 2$\%$\cite{Ba} and 3.5$\%$,\cite{Bac} respectively.
Hydrodynamical calculations, which assume local equilibrium\cite{Kol} agree with central collision data, indicating that thermalization is obtained early in the collision. 
Fig. 18 and 19 show the newest data: $v_2$ vs p$_t$ for a minimum bias trigger.\cite{Du,Sn} Errors shown are statistical only, systematic errors are estimated to be $\sim$ 13 $\%$ at p$_t$=2 GeV/c rising to $\sim$20$\%$ at p$_t$=4.5 GeV/c. The hydrodynamical pattern follows the data at low p$_t$; at higher p$_t$ there is significant discrepancy. Shown in fig.19, calculations which combine a hydrodynamical model at low p$_t$ with a perturbative QCD calculations incorporating partonic energy loss at high p$_t$, describe data quite well. At high p$_t$, elliptic flow is expected to disappear unless there is partonic energy loss in the medium to create the azimuthal anisotropy.\cite{Wa1}

% fig 18 v2 with hydro
\begin{figure}%1
\epsfxsize212pt
\figurebox{}{}{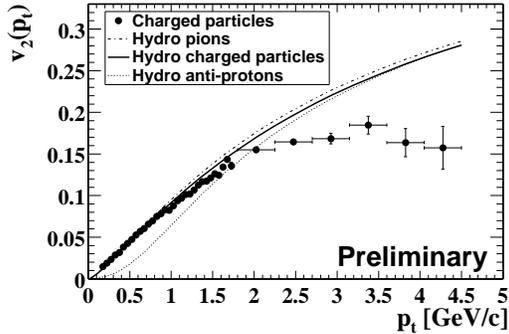}
\caption{ Elliptic flow, $v_2$, for charged particles and minimum bias events, as a function of p$_t$ for Au+Au, compared to hydrodynamical calculations. Errors are statistical only.}
\label{fig:radk}
\end{figure}

% fig 19 v2 vs pt for hydro+qcd
\begin{figure}%1
\epsfxsize212pt
\figurebox{}{}{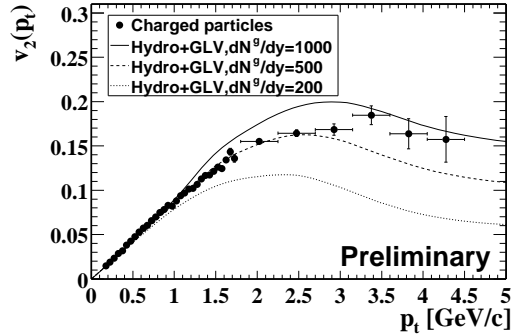}
\caption{Elliptic flow, $v_2$, for charged particles and minimum bias events, as a function of p$_t$ for Au+Au, compared to perturbative QCD calculations. Errors are statistical only.}
\label{fig:radk}
\end{figure}

Further measurements at higher p$_t$ should help elucidate the qualitative agreement seen.

\section{What Next?}%1

Operation of the Relativistic Heavy Ion Collider for physics is just beginning. During the first (2000) and second (2001) runs, the gold beams at energy of $\sqrt{S}$ = 130 and 200 GeV, were accelerated and collided. Analyzed data allowed for significant progress in mapping out the soft physics regime. The global conditions are indeed very different from the ones at SPS (much higher charged particle multiplicity, increased production of antiparticles, low net-baryon density at midrapidity, to name a few). Moreover, the elliptic flow measurements suggest thermalization at an early stage of the collision. In these circumstances, of particular interest are any signals of a new physics, for example ``hard scatterings'', the products of parton scatterings with large momentum transfer resulting from the earliest time during the collision (well before the QGP is expected to form).
 Those scattered partons would subsequently experience the strongly interacting medium and lose energy in hot and dense matter though gluon bremsstrahlung. The most direct consequence of the process, the suppression of high $p_t$ hadron yields (charged and neutral) in central Au+Au collisions in respect to elementary and peripheral collisions, is indeed reported by a number of independent measurements at RHIC. This depletion, if linked unambiguously to ``jet quenching'', can become potentially serious evidence of QGP formation.

RHIC will accelerate ions from protons up to the heaviest nuclei over a range of energies up to 250 GeV for protons and 100 GeV/nucleon for Au nuclei. The RHIC ``jump start'' in 2000 and 2001 from the highest system mass and the top energy, will be followed by systematic studies over the broad range of system masses and energy.

A few years later, the LHC heavy ion data of unprecedentedly high energy (center of mass about a factor 30 greater then RHIC) will begin to compliment the RHIC scientific program. In 2006, the CERN LHC will be the highest energy accelerator operating on Earth. Its approved experimental program includes a strong heavy ion component, with one dedicated heavy ion experiment, ALICE, and an additional heavy ion program in CMS.  

A complete picture of heavy ion collision dynamics at high energies requires the analysis of the complimentary information gained at both RHIC and LHC.

Despite the striking picture emerging from the analyzed RHIC collisions at $\sqrt{S}$=130 and 200 GeV, the highly excited matter produced at RHIC energies is not quite yet at the baryon chemical potential $\mu_B$=0 akin to the Early Universe. The opportunity to discover the fascinating and challenging new science those nuclear collisions present is still ahead.

\section*{Acknowledgments}

This work was supported by the Director, Office of Energy Research, Office of High Energy and Nuclear Physics, Division of Nuclear Physics of the US Department of Energy under Contract DE-AC03-76SF00098.

\end{document}